\documentclass[12pt]{article}

\usepackage{amsmath}
\usepackage{amssymb}
\usepackage{graphicx}

\begin{document}

\title{ \bf ABC-formula and R-operation
for \\ decay processes
}
\author{Dmitrii V. Prokhorenko $~~~~$\\
$~~~~~~~~~$\\
Steklov Mathematical Institute\\
Russian Academy of Sciences\\
Gubkin St. 8, 119991, Moscow, Russia\\
email:prokhor@mi.ras.ru}

\date{}

\maketitle
\begin{abstract}
The vacuum expectation value of the evolution operator for a
general class of Hamiltonians used in quantum field theory
and statistical physics and which include unstable particles is considered.
An exact formula which describes the large time behavior of the
evolution operator is proved.
\end{abstract}
\newpage
\section{Introduction}
The basis object in quantum field theory is scattering matrix, which
describe behavior of the system at infinite intervals of times
\cite{BS,Jos}. But there exists a lot of interesting problems where
one is interested for behavior of the system at large but finite
intervals of times. The general method for studying such problems is
the method of stochastic limit, developed by L. Accardy, Y.G. Lu and
I.V. Volovich \cite{ALV}. The main idea of this method is the
derivation of quantum stochastic differential equations which
describe the evolution operator for small coupling constant
\(\lambda\) and large time \(t\). Higher order corrections to
answers obtained by method of stochastic limit are found in \cite
{PV}.

 We consider the following Hamiltonian in the Fock space:
\begin{eqnarray}
H=H_0+\lambda V, \nonumber
\end{eqnarray}
where \(H_0\) is a free hamiltonian, and V is an interaction,
which is equal to the sum of Wick monomials with kernels from the
Schwartz space. Using perturbation theory we obtain the following
exact formula for the vacuum expectation value of the evolution
operator:
\begin{eqnarray}
\langle 0| U(t)|0\rangle= e^{At+B+C(t)}, \nonumber
\end{eqnarray}
where \(A\) and \(B\) are constants and \(C(t)\rightarrow 0\) as
\(t\rightarrow\infty\). \newline
  \indent This formula was obtained in \cite{AV} by using wave operators
  and perturbation theory. It is called the ABC formula.
  The wave operators method can be used to
  establish this formula only for non decay Hamiltonians. The
  method of perturbation theory admits to consider even the decay
  case. But in \cite{AV} the expression for \(B\) and \(C\) obtained by
  using perturbation theory requires a modification for the decay case, which
  is obtained in the present paper.

  Note that to establish that the integrals, which represent the
  coefficients \(A\) and \(B\) are exist we use the technique
  similar to the technique used in the renormalization theory \cite{BS,Hep,Zav,Smi,KTV}.

  The asymptotics of the evolution operator was studied in
  \cite{Fad,Kul,Hep}

  In the kinetic theory of gases for the determination of the
  transport coefficients is used the density expansion. It was discovered have
   that this expansion contains divergences
  \cite{DC,Dor}.
  These divergences appear when one studies dynamics of finite set of particles
  at large intervals of time. It is possible that the technique of
  renormalization theory can be applied for studying these
  divergences.

  The scattering problem for elementary particles is an important
  problem, because most of elementary particles are unstable \cite{Oku}.
  The stable particles form an irreducible unitary
  representation of the Poincar\(\rm \acute{e} \mit \) group. The
  representation of the Poincar\(\rm \acute{e} \mit \) group
  corresponding to the unstable particles was considered in \cite{U1}.
  A new wave-function renormalization prescription for an unstable
  particle based on the complex-pole mass renormalization was
  suggested in \cite{U2}. There was also proved that conventional
  wave-function renormalization prescription leads to gauge dependent results.

  The Fermi golden rule and large time exponential behavior for
  Lee models was studied in \cite{U3}. About scattering problem for Lee
 models see also \cite{Ar}.

 In \cite{U4} it was suggested that one  can consider unstable particles as
  open quantum systems.

 It is possible that the \(ABC\)-formula could be used to prove the existence of
 wave operators for the cases when one has decay.

 The paper organized as follows. In Section 2 we state the
assumptions on the Hamiltonian \(H\) and formulate the main theorem.
In Section 3 we find sufficient conditions for existence of the
coefficient \(A\). In Section 4 we find the sufficient conditions
for the existence of coefficient \(B\).
\section{The main theorem}
Let \(\mathcal{F}={L^2(\mathbf{R}^d)}\) be the Boson Fock space
over one particle Hilbert space \(L^2(\mathbf{R}^d)\). Let
\(a^+(k)\) and \(a(k)\) are creation and annihilation operator in
\(\mathcal{F}\)
\begin{eqnarray}
{[a(k),a(k)]}={[a^+(k),a^+(k)]}=0,\nonumber\\
{[a(k),a^+(k)]}=0.
\end{eqnarray}
\indent Vacuum state \(|0\rangle\) in \(\mathcal{F}\) is defined by
the relation \(a(k)|0\rangle=0\).
 Consider the following Hamiltonian in the
 Fock space \(\mathcal{F}\)
\begin{eqnarray}
H=\int \omega(k)a^+(k)a(k)dk+\nonumber\\
 \lambda \sum
\limits_{0\leq m,n\leq N} \int v_{m,n}(p_1,...,p_m|p_1,...,p_n)
\prod \limits_{i=1}^m a^+(p_i)dp_i\prod \limits_{j=1}^n a(q_j)dq_j
\end{eqnarray}
\( k,\,p_i,\,q_j \in \mathbf{R}^d\),  \(N=1,2,3...\)\newline
 \(v_{m,n}\) --- are the functions from the Schwartz space of test functions .\newline
 \(v_{0,0}=0,\)\newline
 \indent We consider the following dispersion law
 \begin{eqnarray}
 \omega(k)=\frac{k^2}{2}-\omega_0,\nonumber\\
 \omega_0 \in \mathbf{R},\;\omega_0\neq0. \nonumber
 \end{eqnarray}
 but main results could be generalized to the relativistic case.
 \indent We will study  the evolution operator
 \begin{eqnarray}
 U(t):=e^{itH_0}e^{-itH}
 \end{eqnarray}
 \(t \in \mathbf{R}\).\newline
\indent Friedrichs graph \cite{Hep} by definition is a quadruple
\((V',R,f^+,f^{-})\) where \newline
 \indent \(V'\) is a finite ordered set, called the set of
 vertices,\newline
 \indent \(R\) is a finite set, called the set of lines. \newline
 \indent \(f^+\) and \(f^-\) are the maps
 \begin{eqnarray}
 f^\pm: R\rightarrow V'
 \end{eqnarray}
 such that \(\forall r \in R \; f^+(r) >f^-(r)\).\newline
\indent We will denote by \(V\) the set of all vertices except the
minimal
 one.

 The Friedrichs graph is called a connected graph if for all two
vertices \(v\neq v'\) there exists
 a sequence of vertices and lines \(v=v_0,r_1,v_1,...r_n, v_n=v'\)
 such that for all \(i=1,...,n\)
 \(v_{i-1}=f^-(r), v_i=f^+(r)\) or \(v_{i-1}=f^+(r), v_i=f^-(r)\),
 \newline
 According to the "linked cluster theorem" \(\cite{Hep}\) we have
\begin{eqnarray}
U(t)=:e^{U_c(t)}:.
\end{eqnarray}
Here the index \(c\) in \(U_c(t)\) indicate that one takes only the
connected graphs, and \(:\;:\) means normal order
\(\cite{Hep}\).\newline
 \indent Now we state our main result. \newline
 \indent \textbf{Theorem 1.}
\textit{If \(d\geq3\), then
\begin{eqnarray}
\langle0|e^{-itH}|0\rangle=e^{A t+B+C(t)}
\end{eqnarray}
in the sense of formal power series in \(\lambda\). Here
\(A,B,C(t)\)
--- formal power series in \(\lambda\)
\begin{eqnarray}
A=\sum \limits_{n=2}^\infty A_n\lambda^n, \nonumber\\
B=\sum \limits_{n=2}^\infty B_n\lambda^n, \nonumber\\
C(t)=\sum \limits_{n=2}^\infty C_n(t)\lambda^n
\end{eqnarray}
 and \(C_n(t)\rightarrow 0\)
as \(t\rightarrow\infty\).}\newline

\indent \textbf{ Beginning of the proof.}
  We have
 \begin{eqnarray}
\langle0|e^{-itH}|0\rangle=\langle0|e^{itH_0}e^{-itH}|0\rangle\nonumber
=e^{\langle0|U_c(t) |0\rangle}
\end{eqnarray}
Here
\begin{equation}
\langle0|U_c(t)|0\rangle=\sum \limits_{n=1}^{\infty}(-i\lambda)^n
F_n (t),
\end{equation}
where \(F_n (t)\) --- is the sum over all connected Friedrichs
graphs of the expressions of the form
\begin{eqnarray}
F_n^\Phi (t)= \int \limits_{0}^t dt_n...\int \limits_{0}^{t_1}
dt_0 \int  e^{i(E_nt_n+...+E_0 t_0)} f(p) dp.
\end{eqnarray}
Here f(p) is a product of the kernels of monomials and \(E_i=\sum
\limits_{f^+(r)=v}\omega(p_r)-\sum \limits_{f^-(r)=v}\omega(q_r)\),
and \(\{p_r\}\) are the moments assigned to the lines \(\{r\mid
f^+(r)=v\}\), and \(\{p_r\}\) are moments assigned to the lines
\(\{r\mid f^-(r)=v\}\).\newline
 Let us transform \(F_n^\Phi\). We have
 \begin{eqnarray}
 F_n^\Phi (t)= \int \limits_{0}^t dt_n \int \limits_{0}^{t_n} dt_{n-1}...\int \limits_{0}^{t_1}
dt_0
 \int e^{i(E_nt_n+...+E_0 t_0)} f(p) dp\nonumber\\
 =\int \limits_{0}^{t} dt_0 \int \limits_{0}^{t-t_0}
dt_{n}...\int \limits_{0}^{t_2} dt_1
 \int e^{i(E_nt_n+...+E_1 t_1)} f(p)dp\nonumber\\
=\int \limits_{0}^{t} dt_0 \int \limits_{0}^{t_0} dt_{n}...\int
\limits_{0}^{t_2} dt_1
 \int e^{i(E_nt_n+...+E_1 t_1)} f(p)dp.
  \end{eqnarray}
  Suppose that \(T>t\). Then
\begin{eqnarray}
 F_n^\Phi (t)=\int \limits_{0}^{t} dt_0 \int \limits_{0}^{T} dt_{n}...\int
\limits_{0}^{t_2} dt_1
 \int e^{i(E_nt_n+...+E_1 t_1)} f(p)dp\nonumber\\
-\int \limits_{0}^{t} dt_0 \int \limits_{t_0}^{T} dt_{n}...\int
\limits_{0}^{t_2} dt_1
 \int e^{i(E_nt_n+...+E_1 t_1)} f(p)dp.
\end{eqnarray}
The function
\begin{eqnarray}
 \int e^{i(E_nt_n+...+E_1 t_1)} f(p)dp
\end{eqnarray}
is a continuous function. Therefore it is locally integrable. Let
us suppose that the following limits exist
\begin{eqnarray}
\lim_{T\rightarrow\infty}  \int \limits_{0}^{T} dt_{n}...\int
\limits_{0}^{t_2} dt_1
 \int e^{i(E_nt_n+...+E_1 t_1)} f(p)dp\\
 \lim_{ t \rightarrow \infty}\lim_{ T \rightarrow \infty} \int \limits_{0}^{t} dt_0 \int
\limits_{t_0}^{T} dt_{n}...\int \limits_{0}^{t_2} dt_1
 \int e^{i(E_nt_n+...+E_1 t_1)} f(p)dp
\end{eqnarray}
Denote by  \(A_n\) the sum of expressions (14) over all connected
graphs of degree \(n\), and by \(B_n\) the sum of expressions
(15). We have
\begin{eqnarray}
F_n(t)=A_n t+B_n+C_n(t)
\end{eqnarray}
Here \(C_n(t)\rightarrow0\) as \(t\rightarrow\infty\). We will
prove below that the limits (14), (15) exist if \(d\geq3\).
\section{Existence of coefficients \(A_n\)}
Let us fix some Friedrichs graph \(\Phi\). We have
 \begin{eqnarray}
 I:= A_n^\Phi= \int \limits_{0}^{T} dt_{n}...\int
\limits_{0}^{t_2} dt_1
 \int e^{i(E_nt_n+...+E_1 t_1)} f(p)dp
 \end{eqnarray}
The graph \(\Phi=(V',R,f^+,f^-)\) in (17) is a connected graph.
 \newline
 \indent We will denote by \(V\) the set of all vertices except the minimal
 one. An arbitrary element of \(V\) we will denote by \(v,v \in V\).
So the set \(V\) we can identify with the set
  \(\{1,...,n\}\). \newline
\indent For every \(r \in R\) let us define the following set
\begin{eqnarray}
V_r:=\{v|f^+(r)\geq v>v-1\geq f^-(r)\}.
\end{eqnarray}
For every \(v\in V\) let us define the following set \(R_v\)
\begin{eqnarray}
R_v:=\{r|f+(r)\geq v>v-1\geq f-(r)\}.
\end{eqnarray}
For every \(A\subset V\) put
\begin{eqnarray}
R_A=\{r|\exists v \in A:\;r \in R_v\}
\end{eqnarray}
 It is clear that \(v \in V_r
\Leftrightarrow r \in R_v\)\newline Let us introduce new variables
\(\tau_i:= t_i-t_{i-1}\)\(i=1,...,n\). We have
\begin{eqnarray}
I=\int \limits_0^\infty d\tau_n...\int \limits_0^\infty d\tau_n
 e^{i\omega_0 \sum \limits_{r \in R} \sum \limits_{v \in V_r}
 \tau_v}\times \nonumber\\
 \int \prod \limits_{r} dp_r  e^{-\frac{i}{2} \sum \limits_{r \in R}p_r^2
 \sum \limits_{v \in V_r}
 \tau_v}f(...p_r...)
\end{eqnarray}
At first we suppose that
\begin{eqnarray}
f(...p_r...)= e^{- \frac{1}{2}\sum \limits_{r \in R}p_r^2}
\end{eqnarray}
Then, we have the following expression for \(I\)
\begin{eqnarray}
I=\int \limits_0^\infty d\tau_n...\int \limits_0^\infty d\tau_1
 e^{i\omega_0 \sum \limits_{r \in R} \sum \limits_{v \in V_r}
 \tau_v}\times \nonumber\\
  \prod \limits_{r} \{ (2\pi)^{\frac{d}{2}}
 (1+i\sum \limits_{v \in V_r}
 \tau_v)^{-\frac{d}{2}}\}
\end{eqnarray}
 We want to know when the integral \(I\) exists.\newline
 Suppose \( A \subset V\), and \(\mathcal{O}_A\) is a subset of
 \(\mathbf{R}_+^n\) defined as follows
\begin{eqnarray}
\mathcal{O}_A=\{(\tau_1,...,\tau_n)|\;\tau_i<1 \;\rm if \mit \;i
\in A;\; \tau_i\geq1 \;\rm if \mit \;i \notin A\}
\end{eqnarray}
The Integral \(I\) exists if and only if \(I\) exists over each
domain \(\mathcal{O}_A\). The integral \(I\) over domain
\(\mathcal{O}_\emptyset\) exists if the following integral exists
\begin{eqnarray}
J=\int \limits_1^\infty d\tau_n...\int \limits_1^\infty d\tau_1
  \prod \limits_{r}
 \frac{1}{(\sum \limits_{v \in V_r}
 \tau_v)^{\frac{d}{2}}}
\end{eqnarray}
Let us use the Holder inequalities to estimate this integral.
\begin{eqnarray}
| \int d\tau \prod \limits_{k=1}^{n}f_k(\tau)|\leq\prod
\limits_{k=1}^{n}(\int d\tau|f_k(\tau)|^{{q_k}})^{\frac{1}{q_k}}\nonumber\\
0<q_k \leq\infty,\; \sum \limits_{k=1}^{n} \frac{1}{q_k}=1
\end{eqnarray}
The existence of left hand side follows from the existence of
right hand side. After integrating over \(\tau_1\) we have
\begin{eqnarray}
J\leq C_1\int \limits_1^\infty d\tau_n...\int \limits_1^\infty
d\tau_2
  \prod \limits_{r}
 \frac{1}{(\sum \limits_{v \in V_r;v\neq1}
 \tau_v)^{\frac{d}{2}-(q_r^1)^{-1}}}
\end{eqnarray}
By definition \(q_r^1=\infty\) if \(r\notin R_1\), \(\sum
\limits_{R} \frac{1}{q_r}=1\).\newline \indent The integral over
\(\tau_1\) exists if \(\exists
q_r^1:\;\frac{d}{2}-(q_r^1)^{-1}>0\).\newline Let us now integrate
over \(\tau_2\). We have
\begin{eqnarray}
J\leq C_2\int \limits_1^\infty d\tau_n...\int \limits_1^\infty
d\tau_3
  \prod \limits_{r}
 \frac{1}{(\sum \limits_{v \in V_r;v\neq1,2}
 \tau_v)^{\frac{d}{2}-(q_r^1)^{-1}-(q_r^2)^{-1}}}
\end{eqnarray}
The integral over \(\tau_2\) exists if
\(\frac{d}{2}-(q_r^1)^{-1}-(q_r^2)^{-1}>0\). Note that this
condition implies the following condition:
\(\frac{d}{2}-(q_r^1)^{-1}>0\).\newline \indent So by induction we
see that \(J\) exists if the following condition A) is
satisfied.\newline
 A) There exist real numbers \(0\leq s_r^v<\infty\) such that
 \begin{equation}
 \sum \limits_{r \in R_v}s_r^v=1; \sum \limits_{v \in V_r} s_r^v
 <\frac{d}{2}.
 \end{equation}
 This condition is equal to the following condition.\newline
 B) There exists real numbers \(0\leq s_r^v<\infty\) such that
\begin{eqnarray}
\sum \limits_{r \in R_v}s_r^v>1;\\
 \sum \limits_{v \in
V_r}s_r^v
 =\frac{d}{2}.
 \end{eqnarray}
 We only prove that condition B) implies A). Suppose that B) holds.
 Diminish a bit all nonzero \(s_{r_v}\). We will have strong
 inequality in (31), and inequality (30) will be satisfied.
 Then taking all vertices from \(v=1\) to \(v=n\) and diminishing
 \(s_r^v\) we will have equality in (31).\newline

 \indent Let us introduce the notion of quotient-graph. We can receive quotient-graph \(\Phi_A\) \(A \subset
 V\)from \(\Phi\) by action of operation \(T_i\).
 \(T_i\Phi\) by definition is graph, with the set of vertices
 \(T_i V= V-\{i\}\), and the set of lines \( T_i R =\{r|r_+\neq i \bigvee r_-\neq i-1\}\)\newline
 Put by definition
\begin{eqnarray}
\Phi_A=\{\prod \limits_{i \in V \setminus A} T_i\} \Phi
\end{eqnarray}

\indent By definition, the power of divergence of \(\Phi\) is the
number
\begin{eqnarray}
C_\Phi=|V|-\frac{d}{2}|R|.
\end{eqnarray}
\indent Our aim is to find the conditions necessary and sufficient
for validity the condition B). If condition B) is satisfied the
power of divergence of graph and all its quotient-graph is
negative. Confirm this statement only for graph.
\begin{eqnarray}
\frac{d}{2}|R|=\sum \limits_{r \in R} \sum \limits_{v \in V_r}
s_r^v=\sum \limits_{v \in V} \sum \limits_{r \in R_v} s_r^v> \sum
\limits_{v \in V}1=|V|
\end{eqnarray}
So
\begin{eqnarray}
|V|-\frac{d}{2}|R|<0.
\end{eqnarray}
Conversely, if power of divergence of graph and all its quotient
graph is negative then condition B) is satisfied.\newline \indent
Let us denote by \({C}\) the set of coefficient \(s_r^v\)
satisfying the condition (31) and condition: \(s_r^v=0\) if \(r
\notin R_v\). Denote by \(\mathbf{C}\) the set of all elements
\({C}\). We can identify \(\mathbf{C}\) with the subset in
\(\mathbf{R}^N\), where
 \(N=\sum \limits_{v \in V} \sum \limits_{r \in R_v} 1\). This set is convex set i.e. if \(C_1
 \in\mathbf{C}\) and \(C_2
 \in\mathbf{C}\) then \(m_1C_1+m_2C_2\in \mathbf{C}\); \(m_1\geq0\),
 \(m_2\geq0\),
 \(m_1+m_2=1\)\newline
\indent Let us introduce the following notations
\begin{eqnarray}
C_v:= \sum \limits_{r \in R_v} s_r^v
 \end{eqnarray}
\begin{eqnarray}
|C|:=\sum \limits_{v \in V}C_v=\frac{d}{2}|R|
\end{eqnarray}
\begin{eqnarray}
C_A:=\sum \limits_{v \in A}C_v
\end{eqnarray}
Suppose that there exists \(C^k\) \(k=1,...,n\) such that
\(C^k_v>1\) if \(k\neq v\)\newline
 Let us see at \(C^1_1\). If
\(C_1^1>1\) the statement is proved. If
 \( C_1^1\leq1 \), consider the following convex linear
 combination
  \(m_1C^1+m_2C^2\), such that
\begin{eqnarray}
|m_1C^1+m_2C^2|_{\{1\}}>1
\end{eqnarray}
\begin{eqnarray}
|m_1C^1+m_2C^2|_{V\setminus\{1\}}>\frac{d}{2}|R|-1-\varepsilon
\end{eqnarray}
In the same way we can find \(\tilde{C}^k\) \(k=2,...n\) such that
\begin{eqnarray}
|\tilde{C}^k|_{V\setminus\{1\}}>\frac{d}{2}|R|-1-\varepsilon\nonumber\\
\tilde{C}^k_l>1, k\neq l.
\end{eqnarray}
If there exists \(k\) such, that \(\tilde{C}^k_k>1\) the statement
is proved. Conversely, we can construct the sets
\(\tilde{\tilde{C}}^k\), \(k=2,3,...,n\) such, that
\(\tilde{\tilde{C}}^k_l>1\) if \(k\neq l\) and
\begin{eqnarray}
\mid\tilde{\tilde{C}}\mid^k_{V\setminus\{1,2\}}>\frac{d}{2}|R|-2-\varepsilon
\end{eqnarray}
This inductive procedure will break or, we will construct the set
\(\hat{C}\) such that
\begin{eqnarray}
\hat{C}_k>1; k\neq n
\end{eqnarray}
and
\begin{eqnarray}
\hat{C}_n>\frac{d}{2}|R|-V+1-\varepsilon>1
\end{eqnarray}
if \(\varepsilon\) enough  small.\newline
 So  it is necessary to construct \(C^k\). In the same way as
 above we find , what it is enough to construct the sets
\(C^{k,l}\) such that
 \(C^{k,l}_v>1\) if \(v\neq k,l\),
 \(\mid C^{k,l}\mid\geq\frac{d}{2}|R|\) (automatically satisfied)
 and
 \(\mid
 C^{k,l}\mid_{V\setminus\{k\}}>\frac{d}{2}|R|_{V\setminus\{k\}}\).
To construct \(C^{k,l}\) as above it is sufficient to construct
the sets \(C^{k,l,m}\) such that \(C^{k,l,m}_v>1\) if \(v\neq
k,l,m\) and \(\mid
C^{k,l,m}\mid_{V\setminus\{k\}}>\frac{d}{2}|R|_{V\setminus\{k\}}\)
and \(\mid
C^{k,l,m}\mid_{V\setminus\{k,l\}}>\frac{d}{2}|R|_{V\setminus\{k,l\}}\)\newline
So by induction we see, that it is sufficient to each permutation
\(P\) to construct the set \(C^P \in \mathbf{C}\) such that
\begin{eqnarray}
|C^P|_{\{P(1),...,P(k)\}}=\frac{d}{2}|R|_{\{P(1),...,P(k)\}}
\end{eqnarray}
It is easy to see that we can do it by induction go from
quotient-graph \(\Phi_{\{P(1),...,P(k)\}}\) to quotient-graph
\(\Phi_{\{P(1),...,P(k+1)\}}\). Indeed
\(\Phi_{\{P(1),...,P(k)\}}\) is the quotient-graph of graph
\(\Phi_{\{P(1),...,P(k+1)\}}\). Suppose that
for\(\Phi_{\{P(1),...,P(k)\}}\) such set of coefficients is
constructed. Let us construct \(s^{P(k+1)}_r\).\newline
 If \(r \notin R_{P(k+1)}\) we put \(s^{P(k+1)}_r=0\).\newline
If \(r \notin R_{P(k+1)}\) and if \(r \in R_{\{P(1),...,P(k)\}}\)
we put \(s^P{(k+1)}_r=0\).\newline If \(r \in R_{P(k+1)}\) and If
\(r \notin R_{\{P(1),...,P(k)\}}\) we put
\(s^{P(k+1)}_r=\frac{d}{2}\). Then\newline
\begin{eqnarray}
\sum \limits_{v \in \{P(1),...,P(k+1)\}} \sum \limits_{r \in R_v}
s^v_r= \nonumber\\
=\sum \limits_{v \in \{P(1),...,P(k)\}} \sum \limits_{r \in R_v}
s^v_r+ \sum \limits_{r \in R_{P(k+1)}}
s^{P(k+1)}_r=\nonumber\\
=\frac{d}{2}|R_{\{P(1),...,P(k)\}}|+(\frac{d}{2}|R_{\{P(1),...,P(k+1)\}}|-
\frac{d}{2}|R_{\{P(1),...,P(k)\}}|)=\nonumber\\
\frac{d}{2}|R_{\{P(1),...,P(k+1)\}}|.
\end{eqnarray}
So we have prove that \( I\) over \(O_\emptyset\) exists if the
power of divergence of \(\Phi\) and of all its quotient-graphs is
negative. Analysis of convergence of \(I\) over others regions
\(O_A\) \(A\neq\emptyset\) is similar to above. It reduces to the
analysis of convergence of \(J\) for \(\Phi_{V\setminus A}\). So
we have proved
\newline
\indent \textbf{Theorem II.}\textit{ If power of divergence of graph
and all its quotient-graph is negative then \(I\)
 exists.}
 \newline
 \indent \textbf{Theorem III.} \textit{If \(d\geq 3\) then all conditions of theorem II are
 satisfied.}\newline
\indent \textbf{Proof.} It is enough to prove that \(\mid
R\mid\geq\mid V\mid\) for each connected Friedrichs graph. To the
minimal vertex (vertex number 0) there exists the vertex number
\(i_1\) and the line which connect these vertices, in another case
the vertex number 0 is a non-trivial connected component of
\(\Phi\). To the vertices with numbers \(0,\;i_1\) there exists a
vertexes number \(i_2\) and the line which connect this vertex with
\(0\) or \(i_1\), in other case the vertexes \(0,i_1\) and the lines
which ends in \(0\) and \(i_1\) consists non-trivial connected
component, e.c.t. Therefore the number of lines is greater or equal
then number of vertices.\newline \indent \textbf{Theorem IV.}
\textit{Theorem II is valid for an arbitrary function from the
Schwartz space .}\newline \indent \textbf{Proof.}
 \begin{eqnarray}
I=\int \limits_0^\infty d\tau_n...\int \limits_0^\infty d\tau_n
 e^{i\omega_0 \sum \limits_{r \in R} \sum \limits_{v \in V}
 \tau_v}\times \nonumber\\
 \int \prod \limits_{r} dp_r  e^{-\frac{i}{2} \sum \limits_{r \in R}p_r^2
 \sum \limits_{v \in V}
 \tau_v}f(...p_r...)
\end{eqnarray}
This integral can be represented as a sum \(I=\sum
\limits_{V\subset A} I_A\), here \(I_A\) has the same integrand as
\(I\) but this integral is taken over \(\mathcal{O}_A\). Let us
estimate integrand in \(\mathcal{O}_A\). Let \(F\) means Fourier
transform on variables  \(p_r\), \(r \in R_B\), \(B=V\setminus
A\). So we have:
\begin{eqnarray}
| \int \prod \limits_{r} dp_r  e^{-\frac{i}{2} \sum \limits_{r \in
R}p_r^2
 \sum \limits_{v \in V}
 \tau_v}F(F^{-1}(f))(...p_r...)|=\nonumber\\
|\int \prod \limits_{r} dp_r F(e^{-\frac{i}{2} \sum \limits_{r \in
R_B} p_r^2
 \sum \limits_{v \in V}
 \tau_v })e^{-\frac{i}{2} \sum \limits_{r \notin
R_B}p_r^2
 \sum \limits_{v \in V}
 \tau_v} (F^{-1}(f(...p_r...)))|=\nonumber\\
 =|\prod \limits_{r \in R_B} (2\pi)^{\frac{d}{2}}\frac{1}{i(\sum_{v
 \in V_r} \tau_v)}\times\nonumber\\
 \times\int \prod \limits_{r} dp_r (\rm exp \mit ({+\frac{i}{2}\frac{ \sum \limits_{r \in
R_B} p_r^2}{
 \sum \limits_{v \in V}
 \tau_v }}))e^{-\frac{i}{2} \sum \limits_{r \notin
R_B}p_r^2
 \sum \limits_{v \in V}
 \tau_v} (F^{-1}(f(...p_r...)))|\leq\nonumber\\
 \rm const \mit \prod \limits_{r \in R_B} (2\pi)^{\frac{d}{2}}\frac{1}{(\sum_{v
 \in V_r^B} \tau_v)}
\end{eqnarray}
Here \(V_r^B:= V_r\cap B\). This estimate is enough to prove
theorem II.
\section{Existence of expression for B}
Let us fix some Friedrichs graph \(\Phi\). Then
\begin{eqnarray}
L:= B_n^\Phi= \lim_{ t \rightarrow \infty}\lim_{ T \rightarrow
\infty} \int \limits_{0}^{t} dt_0 \int
\limits_{t_0}^{T} dt_{n}...\int \limits_{0}^{t_2} dt_1\times\nonumber\\
\times \int e^{i(E_nt_n+...+E_1 t_1)} f(p)dp=\nonumber\\
 \lim_{ t \rightarrow \infty}\lim_{ T
\rightarrow \infty} \int
\limits_{\tau_1+...+\tau_n<T;\tau_i>0}\prod \limits_{i=1}^{n} d
\tau^i
\rm min \mit(t,\tau_1+...+\tau_n)\times\nonumber\\
e^{i\omega_0 \sum \limits_{r \in R} \sum \limits_{v \in V}
 \tau_v}\times
 \int \prod \limits_{r} dp_r  e^{-\frac{i}{2} \sum \limits_{r \in R}p_r^2
 \sum \limits_{v \in V}
 \tau_v}f(...p_r...).
\end{eqnarray}
Our aim is to prove that this integral converges if conditions of
theorem II are satisfied. We can represent this integral as a sum
\newline \(L=\sum \limits_{A\subset V} L_A\), here \(L_A\) has the
same integrand as \(L\), but taken over the region
\(\mathcal{O}_A\). For the sake of simplicity we consider only
\(L_\emptyset\). In the fist time we transpose limits in
\(L_\emptyset\), prove the convergence of obtained integral, and
prove that the transposition of the limits does not change the
value of integral. If we transpose the limits we will have
\begin{eqnarray}
K_\emptyset(T)=\int \limits_{\tau_1+...+\tau_n<T;\tau_i>1}
\prod_{i=1}^n d\tau^i
(\tau_1+...+\tau_n)\times\nonumber\\
e^{i\omega_0 \sum \limits_{r \in R} \sum \limits_{v \in V}
 \tau_v}\times
 \int \prod \limits_{r} dp_r  e^{-\frac{i}{2} \sum \limits_{r \in R}p_r^2
 \sum \limits_{v \in V}
 \tau_v}f(...p_r...)
\end{eqnarray}
or
\begin{eqnarray}
K_\emptyset(T)=\int \limits_{\tau_1+...+\tau_n<T;\tau_i>1}
\prod_{i=1}^n d\tau^i (\tau_1+...+\tau_n)e^{i\omega_0 \sum
\limits_{r \in R} \sum \limits_{v \in V}
 \tau_v}\times\nonumber\\
 \prod \limits_{r}\frac{1}
 {(2\pi)^{\frac{d}{2}}}\frac{1}{(\sum \limits_{v
 \in V_r} \tau_v)}f(\tau_1,...,\tau_n)
\end{eqnarray}
where
\begin{eqnarray}
f(\tau_1,...,\tau_n)=(-i)\int \prod \limits_{r \in R} dx_r
F^{-1}(x_r)e^{-\frac{i}{2}\sum \limits_{r \in R} x^2_r
\frac{1}{\sum \limits_{v
 \in V_r} \tau_v}}
\end{eqnarray}
\indent \textbf{Lemma 1.}
\begin{eqnarray}
|\frac{d}{d\tau}f(\tau,...,\tau,\tau_{k+1},...,\tau_n)|\leq
\frac{\rm const \mit}{\tau}
\end{eqnarray}
\indent \textbf{Proof.}
\begin{eqnarray}
\frac{d}{d\tau}f(\tau,...,\tau,\tau_{k+1},...,\tau_n)=\nonumber\\
 \frac{1}{2}
\prod \int \limits_{r \in R} dx_r F^{-1}(x_r)e^{-\frac{i}{2}\sum
\limits_{r \in R} x^2_r \frac{1}{\sum \limits_{v
 \in V_r} \tau_v}}
 \nonumber\\
\times \sum \limits_{v \in \{1,...,k\}}\sum \limits_{r \in R_v}
\frac{x_r^2}{(\sum \limits_{v
 \in V_r} \tau_v)^2}|_{\tau_1=...=\tau_k=\tau}
 \end{eqnarray}
 But \(\forall v \in \{1,...,k\}\)
\begin{eqnarray}
\sum \limits_{r \in R_v}
\frac{1}{\sum \limits_{v
 \in V_r} \tau_v}|_{\tau_1=...=\tau_k=\tau}\leq \frac{\rm const \mit}{\tau}
 \end{eqnarray}
We can estimate the integral by constant. The lemma is
proved.\newline \indent \textbf{Lemma 2.} \textit{If conditions of
theorem II are satisfied the following limit exists}
\begin{eqnarray}
\lim_{T\rightarrow\infty}
K_\emptyset(T)^k=\lim_{T\rightarrow\infty} \int
\limits_{k\tau_k+...+\tau_n<T;\tau_i>1;\tau_k>...>\tau_n}
\prod_{i=k}^n d\tau_i\times\nonumber\\
(\tau_1+...+\tau_n)e^{i\omega_0 \sum \limits_{r \in R} \sum
\limits_{v \in V}
 \tau_v}\times\nonumber\\
 \prod \limits_{r}\frac{1}
 {(2\pi)^{\frac{d}{2}}}\frac{1}{(\sum \limits_{v
 \in V_r} \tau_v)^{\frac{d}{2}}}f(\tau_1+...+\tau_n)|_{\tau_1=...=\tau_k}
\end{eqnarray}
\indent \textbf{Proof.} In the first time suppose that \(k=n\).
The integral is equal
\begin{eqnarray}
K_\emptyset^n(T)=n\int \limits_{1}^{\frac{T}{n}} d\tau \tau
  e^{i\omega_0 \sum \limits_r |V_r|\tau} \prod \limits_{r}\frac{1}
 {(2\pi)^{\frac{d}{2}}}\frac{1}{(\sum \limits_{v
 \in V_r} \tau)^{\frac{d}{2}}}f(\tau...\tau)=\nonumber\\
 =\frac{n}{i\omega_0 N} \{\frac{T}{n} e^{i\omega_0 \sum \limits_r |V_r|\frac{T}{n}}
\prod \limits_{r} \frac{1}
 {(2\pi)^{\frac{d}{2}}}\frac{1}{(\sum \limits_{v
 \in V_r} \tau)^{\frac{d}{2}}} f(\tau...\tau)|_{\tau=\frac{T}{n}}- \nonumber\\
-e^{i\omega_0 \sum \limits_r |V_r|} \prod \limits_{r}
\frac{1}{(2\pi)^{\frac{d}{2}}} \frac{1}{(|V_r|)^{\frac{d}{2}}} f(1...1)+\nonumber\\
+(\frac{d}{2}|R|-1)\int \limits_{1}^{\frac{T}{n}} d\tau
\frac{1}{\tau^{\frac{d}{2}|R|}}\prod \limits_{r}
\frac{1}{(2\pi)^{\frac{d}{2}}}
\frac{1}{(|V_r|)^{\frac{d}{2}}}e^{i\omega_0 \sum \limits_r
|V_r|\tau}f(\tau...\tau)-\nonumber\\
-\prod \limits_{r} \frac{1}{(2\pi)^{\frac{d}{2}}}
\frac{1}{(|V_r|)^{\frac{d}{2}}}\int \limits_{1}^{\frac{T}{n}}
d\tau\frac{\tau}{\tau^{\frac{d}{2}|R|}}\frac{d}{d\tau}f(\tau,...,\tau)e^{i\omega_0
\sum \limits_r |V_r|\tau}\}
\end{eqnarray}
 But \(\frac{d}{2}|R|>|V|\geq1\) Therefore, the first term tends
 to zero, the second term is constant, the third integral converges
 and fourth converges because
\begin{eqnarray}
|\frac{d}{d\tau}f(\tau,...,\tau)|\leq \frac{\rm const \mit}{\tau}.
\end{eqnarray}
Let us suppose that the statement of lemma has proved for \(k+1\).
Let us prove the statement for \(k\). Using integration by parts
on \(d\tau_{k+1}\) we have
\begin{eqnarray}
K_\emptyset(T)^k=\int
\limits_{k\tau_k+...+\tau_n<T;\tau_i>1;\tau_k>...>\tau_n}
(\tau_1+...+\tau_n)e^{i\omega_0 \sum \limits_{r \in R} \sum
\limits_{v \in V}
 \tau_v}\times\nonumber\\
 \prod \limits_{r}\frac{1}
 {(2\pi)^{\frac{d}{2}}}\frac{1}{(\sum \limits_{v
 \in V_r}
 \tau_v)^{\frac{d}{2}}}f(\tau_1+...+\tau_n)|_{\tau_1=...=\tau_k} \prod \limits_{i=k}^n d\tau_i=\nonumber\\
 =\frac{1}{i\omega_0 N}\{T \int
\limits_{(k+1)\tau_{k+1}+...+\tau_n<T;\tau_i>1;\tau_{k+1}>...>\tau_n}
\prod \limits_{i=k+1}^n d\tau_i\times \nonumber\\ \times\prod
\limits_{r \in R} \frac{1}{(2\pi)^{\frac{d}{2}}}\frac{1}{(\sum
\limits_{v
 \in V_r}
 \tau_v)^{\frac{d}{2}}}e^{i\omega_0 \sum \limits_{r \in R} \sum
\limits_{v \in V}
 \tau_v}f(\tau_1+...+\tau_n)|_{\tau_1=...=\tau_k=
 \frac{T-\tau_{k+1}-....-\tau_n}{k}}-\nonumber\\
- \int
\limits_{(k+1)\tau_{k+1}+...+\tau_n<T;\tau_i>1;\tau_(k+1)>...>\tau_n}
\prod \limits_{i=k+1}^n d\tau_i\times \nonumber\\
 \times\prod
\limits_{r \in R} \frac{1}{(2\pi)^{\frac{d}{2}}}\frac{1}{(\sum
\limits_{v
 \in V_r}
 \tau_v)^{\frac{d}{2}}}e^{i\omega_0 \sum \limits_{r \in R} \sum
\limits_{v \in V}
 \tau_v}(\tau_1+...+\tau_n)f(\tau_1+...+\tau_n)|_{\tau_1=...=\tau_k=\tau_{k+1}}-
\nonumber\\
 -\int
\limits_{k\tau_k+...+\tau_n<T;\tau_i>1;\tau_k>...>\tau_n} \prod
\limits_{i=k}^{n}d\tau_i \frac{d}{d\tau_k}(
\{(k\tau_k+...+\tau_n)f(\tau_1+...+\tau_n)\times \nonumber\\
\prod \limits_{r}\frac{1}
 {(2\pi)^{\frac{d}{2}}}\frac{1}{(\sum
\limits_{v
 \in V_r}
 \tau_v)^{\frac{d}{2}}}\}|_{\tau_1=...=\tau_k})\times e^{i\omega_0 \sum \limits_{r \in R}
  \sum \limits_{v \in V}
 \tau_v}
\end{eqnarray}
Here \(N=\sum \limits_{l=1}^{k}\sum \limits_{r \in R_l}1\). The
second term is equal to \(K_\emptyset^{k+1}(T)\) hence it has a
limit as \(T\rightarrow\infty\). To estimate the first term we
will use Holder inequality. We will take coefficients \(q_r^k\) as
in theorem II. Note that \(\tau_1=...=\tau_k>\frac{T}{n}\). Let us
denote the first term by \(Q\). Then
 \begin{equation}
 \mid Q\mid\leq T \rm const \mit \prod \limits_{r \in R_{\{1,...,k\}}}
 \frac{1}{T^{\{\frac{d}{2}-\sum \limits_{v=k+1}^n (q_r^k)^{-1}\}}}
 \end{equation}
 Because the coefficients \((q_r^k)\) are the same as in theorem II
 we have
 \begin{equation}
\frac{d}{2}-\sum \limits_{v=k+1}^n (q_r^k)^{-1}\geq
(q_r^1)^{-1}+\varepsilon
\end{equation}
For some \(\varepsilon\). Hence
\begin{equation}
\mid Q\mid\leq\frac{1}{T^\varepsilon}\frac{T}{T^{\sum \limits_r
(q_r^1)^{-1}}}=\frac{1}{T^\varepsilon}\rightarrow 0
\end{equation}
So the first term tends to zero. Let us use the Leibnitz rule for
the third term. We get the sum of three term.
\begin{eqnarray}
k\int \limits_{k\tau_k+...+\tau_n<T;\tau_i>1;\tau_k>...>\tau_n}
\prod \limits_{i=k}^{n}d\tau_i e^{i\omega_0 \sum \limits_{r \in R}
\sum \limits_{v \in V}
 \tau_v}\times\nonumber\\
 \prod \limits_{r}\frac{1}
 {(2\pi)^{\frac{d}{2}}}\frac{1}{(\sum \limits_{v
 \in V_r}
 \tau_v)^{\frac{d}{2}}}f(\tau_1+...+\tau_n)|_{\tau_1=...=\tau_k}-\nonumber\\
-\int \limits_{k\tau_k+...+\tau_n<T;\tau_i>1;\tau_k>...>\tau_n}
\prod \limits_{i=k}^n d\tau_i
 \{\sum \limits_{v \in {1,...,k}} \sum \limits_{r \in R_v}
 \frac{1}{\sum \limits_{v \in V_r}\tau_v}\}
 (k\tau_k+...+\tau_n)e^{i\omega_0 \sum \limits_{r \in R} \sum
\limits_{v \in V}
 \tau_v}\times\nonumber\\
 \prod \limits_{r}\frac{1}
 {(2\pi)^{\frac{d}{2}}}\frac{1}{(\sum \limits_{v
 \in V_r}
 \tau_v)^{\frac{d}{2}}}f(\tau_1+...+\tau_n)|_{\tau_1=...=\tau_k}+\nonumber\\
 +\int \limits_{k\tau_k+...+\tau_n<T;\tau_i>1;\tau_k>...>\tau_n}\prod \limits_{i=k}^n d\tau_i
(\tau_1+...+\tau_n)e^{i\omega_0 \sum \limits_{r \in R} \sum
\limits_{v \in V}
 \tau_v}\times\nonumber\\
 \prod \limits_{r}\frac{1}{(\sum \limits_{v
 \in V_r}
 \tau_v)^{\frac{d}{2}}}\frac{d}{d\tau_k}f
 (\tau_k,...,\tau_k,...,\tau_n)|_{\tau_1=...=\tau_k}.
 \end{eqnarray}
 Because \(k\tau_k+...+\tau_n<\tau_k n\) and
 \begin{eqnarray}
 \prod \limits_{r}\frac{1}{(\sum \limits_{v
 \in V_r}
 \tau_v)^{\frac{d}{2}}}\leq\prod \limits_{r\in R_{\{k,...,n\}}}
\frac{1}{(\sum \limits_{v
 \in V_r\cap\{k,...,n\}}\tau_v)^{\frac{d}{2}}}
 \end{eqnarray}
 the Cauchy difference for the last term can be estimated by the Cauchy difference
 for \(J\) constructed by factor graph
 \(\Phi_{\{k,...,n\}}\). So the last term converges. So if the statement of the lemma holds for \(k+1\)
 then it is holds for \(k\). The lemma is proved.\newline
 Note that \(K_\emptyset\) is the sum of the terms which can be
 estimated as \(K_\emptyset(T)^k\).
So we have proved the following lemma. \newline

\indent \textbf{Lemma 3.}\textit{The expression \(K_\emptyset(T)\)
have the limit as \(T \rightarrow \infty\).}
\newline
\indent \textbf{Lemma 4.} \textit{We can transpose the limits in
(49)} .\newline
 \textbf{Proof.} Let us consider the difference between the limit of
 (49)as
 \(T\rightarrow\infty\) and
 \(K_\emptyset(t)\) we have
\begin{eqnarray}
L_\emptyset(t)-K_\emptyset(t)=\nonumber\\
t\int \limits_{\tau_1+...+\tau_n>t;\tau_i>1} \prod \limits_{i=1}^n
d\tau_i e^{i\omega_0 \sum \limits_{r \in R} \sum \limits_{v \in V}
 \tau_v}\times\nonumber\\
 \times\prod \limits_{r}\frac{1}
 {(2\pi)^{\frac{d}{2}}}\frac{1}{(\sum \limits_{v
 \in V_r} \tau_v)^{\frac{d}{2}}}f(\tau_1,...,\tau_n)
 \end{eqnarray}
 Divide the region of integration into the following sectors
\begin{eqnarray}
\mathcal{O}_\sigma=\{(\tau_1,...,\tau_n)|\tau_i>1;\tau_{\sigma(1)}>...\tau_{\sigma(n)}\}
\end{eqnarray}
Here \(\sigma\) --- is an permutation of the set \(\{1,...,n\}\).
Consider only the integral over \(\mathcal{O}_{id}\). We can
consider other integrals in the same way. Now we will integrate by
parts on \(\tau_1\). We have
\begin{eqnarray}
L_\emptyset(T)-K_\emptyset(T)=\nonumber\\
\frac{T}{i\omega_0 N}\{\lim_{T_1\rightarrow\infty} \int
\limits_{T<2\tau_2+...+\tau_n<T_1;\tau_i>1; \tau_2>...>\tau_n}
\prod \limits_{i=2}^n d\tau_i e^{i\omega_0 \sum \limits_{r \in R}
\sum \limits_{v \in V}
 \tau_v}\times\nonumber\\
 \prod \limits_{r}\frac{1}
 {(2\pi)^{\frac{d}{2}}}\frac{1}{(\sum \limits_{v
 \in V_r}
 \tau_v)^{\frac{d}{2}}}f(\tau_1,...,\tau_n)\mid_{\tau_1={T_1-\tau_2-...-\tau_n}}-\nonumber\\
-\lim_{T_1\rightarrow\infty} \int
\limits_{T<2\tau_2+...+\tau_n<T_1;\tau_i>1;\tau_2>...>\tau_n}\prod
\limits_{i=2}^n d\tau_i e^{i\omega_0 \sum \limits_{r \in R} \sum
\limits_{v \in V}
 \tau_v}\times\nonumber\\
 \prod \limits_{r}\frac{1}
 {(2\pi)^{\frac{d}{2}}}\frac{1}{(\sum \limits_{v
 \in V_r}
 \tau_v)^{\frac{d}{2}}}f(\tau_1,...,\tau_n)\mid_{\tau_1=\tau_2}-\nonumber\\
 -\lim_{T_1\rightarrow\infty} \int
\limits_{2\tau_2+...+\tau_n<T;\tau_i>1;\tau_2>...>\tau_n}\prod
\limits_{i=2}^n d\tau_i e^{i\omega_0 \sum \limits_{r \in R} \sum
\limits_{v \in V}
 \tau_v}\times\nonumber\\
 \times\prod \limits_{r}\frac{1}
 {(2\pi)^{\frac{d}{2}}}\frac{1}{(\sum \limits_{v
 \in V_r}
 \tau_v)^{\frac{d}{2}}}f(\tau_1,...,\tau_n)\mid_{\tau_1={T-\tau_2-...-\tau_n}}-\nonumber\\-
  \lim_{T_1\rightarrow\infty} \int
\limits_{T<\tau_1+...+\tau_n<T_1;\tau_i>1;\tau_1>...>\tau_n}\prod
\limits_{i=1}^n d\tau_i e^{i\omega_0 \sum \limits_{r \in R} \sum
\limits_{v \in V}
 \tau_v}\times\nonumber\\
\times \frac{d}{d\tau_1}\{\prod \limits_{r}\frac{1}
 {(2\pi)^{\frac{d}{2}}}\frac{1}{(\sum \limits_{
 \in V_r}
 \tau_v)^{\frac{d}{2}}}f(\tau_1,...,\tau_n)\}\}
\end{eqnarray}
Here \(N=\sum \limits_{r \in R_1}1\). The first term is equal to
zero. The third term tends to zero as  \(T\rightarrow\infty\). The
proof of this fact is the same as the proof for the first term in
(59). The fourth term tends to zero as \(T\rightarrow\infty\). The
proof is similar to the proof for the last term in (59). We can
estimate the second term as \(L_\emptyset(t)-K_\emptyset(t)\) for
a diagram with \(V-1\) vertexes.  The lemma IV is proved.\newline
 These four lemmas implies the following.\newline
\indent \textbf{Theorem V.} \textit{The coefficient \(B_n\) is well
defined and is equal to the sum of the terms of the form
 (if conditions of theorem V are satisfied)
\begin{eqnarray}
\lim_{T\rightarrow\infty}\int
\limits_{\tau_1+...+\tau_n<T;\tau_i>0}
(\tau_1+...+\tau_n)\times\nonumber\\
e^{i\omega_0 \sum \limits_{r \in R} \sum \limits_{v \in V}
 \tau_v}\times
 \int \prod \limits_{r} dp_r  e^{-\frac{i}{2} \sum \limits_{r \in R}p_r^2
 \sum \limits_{v \in V}
 \tau_v}f(...p_r...)
 \end{eqnarray}
 where \(f(...p_r...)\) is the product of monomials of kernels.}\newline
 Therefore the coefficients \(A_n\) and \(B_n\) exist and Theorem I complectly
 proved.\newline
 \indent Note that theorem II is analogous to the theorem about
 convergence Feynman integrals in quantum field theory.
 \section{Conclusion} We have derived the general formula which describes the large time
 behavior of the vacuum expectation value of the evolution operator
 for the general class of
 Hamiltonias describing the decay processes.
\section{Acknowledgements}
 I wold like to thank my supervisor I.V. Volovich for very useful discussions.\newline
 \indent This work was partially supported by the Russian
 Foundation of Basis Reasearch (project 05-01-008884), the grant
 of the president of the Russian Federation (project
 NSh-6705.2006.1) and the program "Modern problems of
 theoretical mathematics" of the Mathematical sciences department
 of the Russian Academy of Sciences.

\end{document}